\documentclass[twocolumn,prb,superscriptaddress,noshowpacs,epsf]{revtex4}

\usepackage{amssymb}
\usepackage{amsmath}
\usepackage{graphicx,bm}
\usepackage{color}
\usepackage{epsfig}

\begin{document}

\title{A simple design of an artificial electromagnetic black hole}

\author{Wanli Lu}
\affiliation{School of Physical Science and Technology, Soochow
University, Suzhou, Jiangsu 215006, China}

\affiliation{Surface Physics Laboratory, Department of Physics,
Fudan University, Shanghai 200433, China}

\author{JunFeng Jin}
\affiliation{Surface Physics Laboratory, Department of Physics,
Fudan University, Shanghai 200433, China}

\author{Huanyang Chen}
\email [To whom correspondence should be addressed.
\\Electronicaddress: ]{kenyon@ust.hk}
\affiliation{School of Physical Science and Technology, Soochow
University, Suzhou, Jiangsu 215006, China}

\author{Zhifang Lin}
\affiliation{Surface Physics Laboratory, Department of Physics,
Fudan University, Shanghai 200433, China}

\date{\today}

\begin{abstract}
We study the properties of an artificial electromagnetic black
hole for transverse magnetic modes rigorously. A multi-layered
structure of such a black hole is then proposed as a reduced
variety for easy experimental realizations. An actual design of
composite materials based on the effective medium theory is given
eventually with only five kinds of real isotropic materials. The
finite element method confirms the functionality of such a simple
design.
\end{abstract}

\pacs{ }

\maketitle

\section{Introduction}

Transformation optics \cite{ulf,pen} is a useful tool to design
many novel wave manipulation devices
\cite{sch,liu,tre,val,gab,smo,chen,mei,ma}. Its general form, the
``general relativity in electrical engineering" \cite{njp}, was
proposed to mimic cosmic optical properties \cite{miao,chy}. A
further method was also suggested \cite{gen}, which can transmute
the anisotropic parameters from the general form into isotropic
ones. An isotropic ``optical black hole" was given as well based
on such a transmuting method \cite{gen}. Another approach to a
broadband absorber was also proposed from Hamiltonian optics
\cite{apl}. Such an absorber can be termed as an effective
`optical black hole" \cite{apl}, which was later implemented by
using nonresonant metamaterial units in microwave frequencies for
transverse electric(TE) modes \cite{cui}. The experiment
demonstrated the importance of metamaterials in implementing the
related transformation media devices once again. However, the
structure is a bit complicated and will be very challenging to
adapt for higher frequencies.

In this paper, we will consider the transverse magnetic(TM) modes
of such an artificial electromagnetic(EM) black hole \cite{cui}
and give a simple design with several kinds of real materials. In
the following, we begin with the rigorous studies of TM modes of
such an EM black hole in section 2. Then we propose a simple
design based on composite materials in section 3. Finally, we give
conclusions in section 4.

\section{The rigorous calculations of an artificial black hole for TM modes}\label{theory}

\begin{figure}[htbp]
  \centering
  \includegraphics[width=5cm]{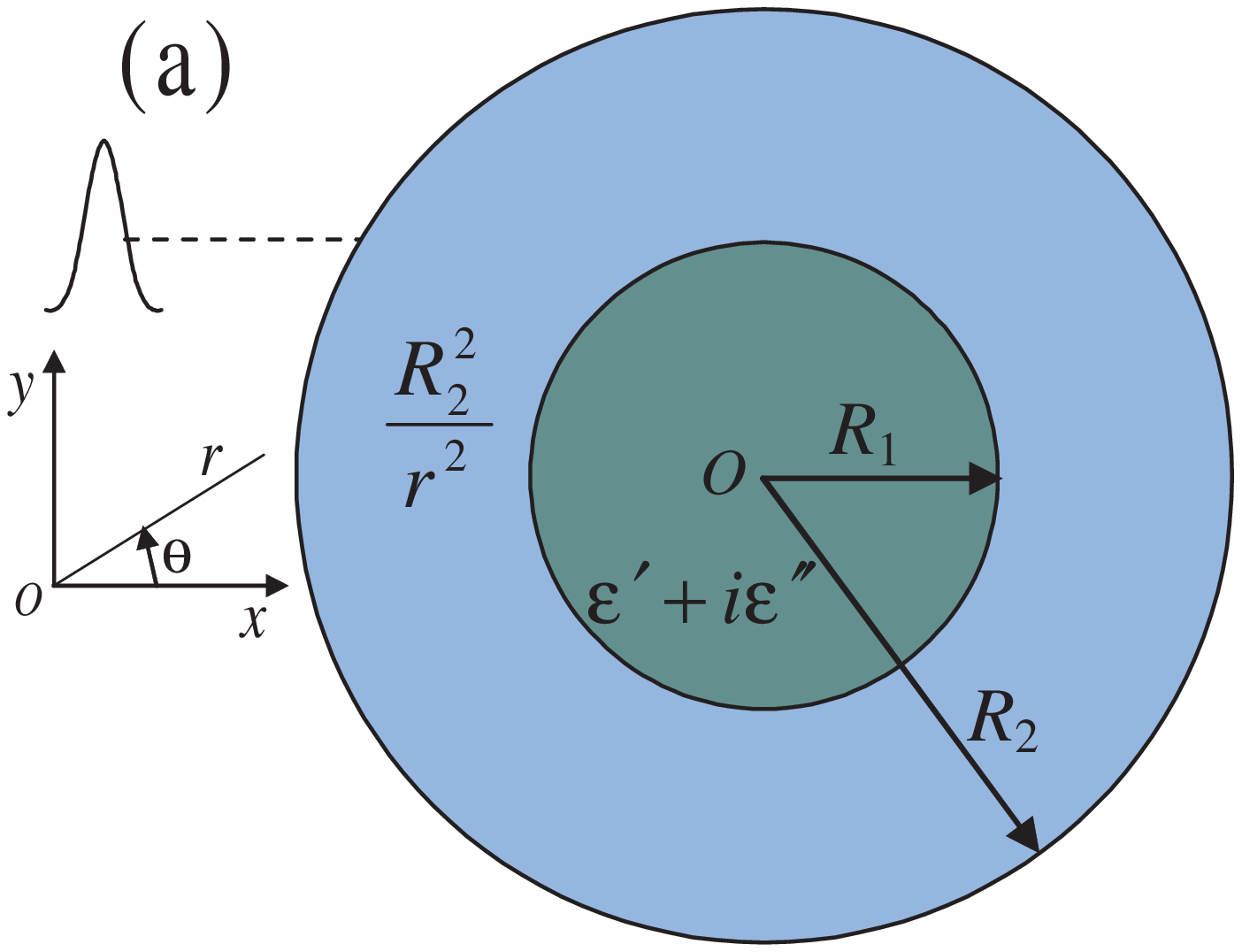}
  \includegraphics[width=6cm]{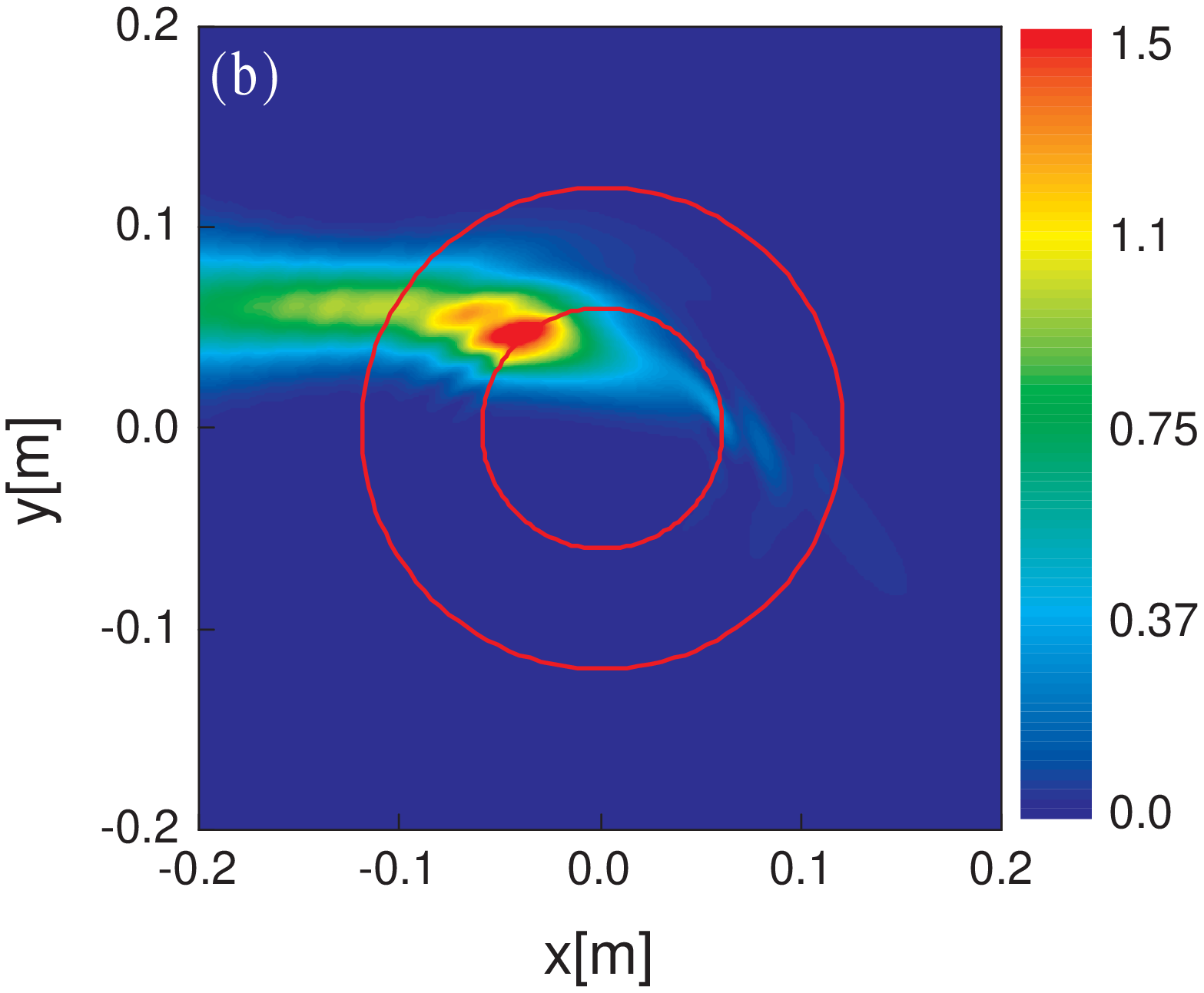}
  \caption{(a). The schematic plot of an incident Gaussian beam interacting with an artificial EM black hole. $R_2$ is the radius of the ``event horizon" described in Ref. \cite{apl}, $R_1$ is the radius of the inner absorbing core.
  (b). The magnetic field intensity pattern for an incident Gaussian beam interacting with the EM black hole, which is calculated based on rigorous solutions Eqs. (\ref{solution})-(\ref{core}).}\label{01}
\end{figure}

We start from the same permittivity profile of the artificial EM
black hole as in Ref. \cite{apl},
\begin{equation}\label{1}
\epsilon(r)=
\left\{%
\begin{array}{ll}
    1, & \hbox{$r>R_2$,} \\
    \frac{R_2^2}{r^2}, & \hbox{$R_1<r<R_2$,} \\
    \epsilon^{\prime}+i\epsilon^{\prime\prime}, & \hbox{$r<R_1$,} \\
\end{array}%
\right.
\end{equation}
where $R_1=R_2\sqrt{\frac{1}{\epsilon^{\prime}}}$ with the
geometry parameters described in Fig.\ref{01}(a).

As it is mentioned in Ref. \cite{apl}, the TE and TM polarizations
decouple and can be solved independently. The TE EM black hole has
been studied rigorously \cite{apl} and implemented with
nonresonant metamaterial units \cite{cui}. With similar
procedures, we will focus on the TM modes with the magnetic field
along the $\hat{z}$-direction and propose a much simpler design in
the following sections.

In the region $R_1<r<R_2$, the general solutions of the magnetic
field in $\hat{z}$-direction can be written as
\cite{JMO,semiclassical theory1},
\begin{equation}\label{in}
    H_z^{in}(r,\theta; t)=\sum^\infty_{n=-\infty}H_z^{in}(r)\exp(in\theta-i\omega t),
\end{equation}
where $n$ is the angular momentum number, $\omega$ is the angular
frequency, and $H_z^{in}(r)$ satisfies,
\begin{equation}\label{function}
    r^2\frac{\text{d}^2H_z^{in}(r)}{\text{d}r^2}+3r\frac{\text{d}H_z^{in}(r)}{\text{d}r}+(k_0^2 R_2^2-n^2)H_z^{in}(r)=0,
\end{equation}
whose solutions can be expressed as
\begin{eqnarray}\label{solution}
    H_z^{in}(r)=A_n
    \left(\frac{r}{R_2}\right)^{-1+\sqrt{n^2-k_0^2R_2^2+1}}\\\nonumber
    +B_n \left(\frac{r}{R_2}\right)^{-1-\sqrt{n^2-k_0^2
    R_2^2+1}}
\end{eqnarray}
with the wave vector of light in vacuum $k_0=\frac{\omega}{c}$.

In the region $r>R_2$, we suppose that an Gaussian beam is
incident with its magnetic field in $\hat{z}$-direction
\cite{semiclassical theory1, wu and guo},
\begin{equation}\label{inc}
    H_z^{inc}(r,\theta; t)=\sum^\infty_{n=-\infty}P_n i^n J_n
(k_0r)\exp(in\theta-i\omega t),
\end{equation}
where $J_n(k_0r)$ are Bessel functions of order $n$, and $P_n$ is
determined by \cite{semiclassical theory1, wu and guo},
\begin{eqnarray}\label{An}
    P_n=\frac{W_0}{2\sqrt{\pi}}\int^\infty_{-\infty}\exp(-\frac{1}{4}k_y^2W_0^2-i\sqrt{k^2-k_y^2}x_0\\\nonumber
    -ik_yy_0-in\alpha)\text{d}k_y,
\end{eqnarray}
where $\lambda$ is the wavelength of the Gaussian beam, $W_0$ is
half of the beam waist, $(x_0, y_0)$ is beam center in Cartesian
coordinates, and $\alpha=\sin^{-1}(\frac{k_y}{k})$ is the angle
between the wave vector and $\hat{x}$-axis. The scattering waves
are assumed to be,
\begin{equation}\label{sca}
    H_z^{sca}(r,\theta; t)=\sum^\infty_{n=-\infty}Q_n i^n H_n^{(1)}(k_0r)\exp(in\theta-i\omega t),
\end{equation}
where $H_n^{(1)}(k_0r)$ are the first kind Hankel functions with
the angular momentum number $n$. Hence the total magnetic field
$H_z^{out}(r,\theta)$ is the summation of the incident field
$H_z^{inc}(r,\theta)$ and the scattering field
$H_z^{sca}(r,\theta)$,
\begin{eqnarray}\label{out}
    H_z^{out}(r,\theta; t)=\sum^\infty_{n=-\infty}i^n[P_n J_n(k_0r)+Q_n H_n^{(1)}(k_0r)]\\\nonumber
    \times\exp(in\theta-i\omega
t).
\end{eqnarray}

In the region $r<R_1$, the absorbing core is isotropic with
permittivity $\epsilon^{\prime}+i\epsilon^{\prime\prime}$, the
magnetic field in $\hat{z}$-direction can be written as,
\begin{equation}\label{core}
    H_z^{core}(r,\theta; t)=\sum^\infty_{n=-\infty}C_nJ_n(k_0\sqrt{\epsilon^{\prime}+i\epsilon^{\prime\prime}}r)\exp(in\theta-i\omega t).
\end{equation}

From the continuous conditions on the boundaries $r=R_1$ and
$r=R_2$, \emph{i.e.}, the continuities of the magnetic field in
$\hat{z}$-direction and its normal derivative, the coefficients
$Q_n$, $A_n$, $B_n$, and $C_n$ can be uniquely determined in terms
of the incident coefficients $P_n$.

In this paper, we will set $R_1=60 mm$, $R_2=120 mm$,
$\epsilon^\prime=4.0$, $\epsilon^{\prime\prime}=0.33$
\cite{inabs}, and $\lambda=30 mm$ for instance. The waist of the
incident Gaussian beam is $60 mm$ and the beam center is at $(-120
mm, 60 mm)$. We use a cut-off angular momentum number
$n_{max}=100$ during the calculations. The magnetic field
intensity pattern $|H_z|^{2}$ is plotted in Fig.\ref{01}(b), which
shows that the beam is bent enormously toward and absorbed by the
inner core. Thereby the above absorbing system can be termed as an
``EM black hole" \cite{apl}.

The absorption cross section per unit length can be written as
\cite{bohren},
\begin{equation}\label{acs}
    c_{abs}=-\frac{4}{k_0}\sum^\infty_{n=-\infty}(|D_n|^2+\text{Re}\{D_n\}),
\end{equation}
with $D_n=Q_n/P_n$.  Substituting the above parameters, the
absorption cross section of the present artificial black hole is
0.2231, which is close to its geometry cross section $2R_2$
\cite{bohren}. That means this artificial black hole can sever as
a nearly perfect absorber as in Ref. \cite{apl}.

\section{Two-step approach to a simpler design of an EM black hole, a layered structure and an actual design}

\begin{figure}[htbp]
  \centering
  \includegraphics[width=0.45\textwidth]{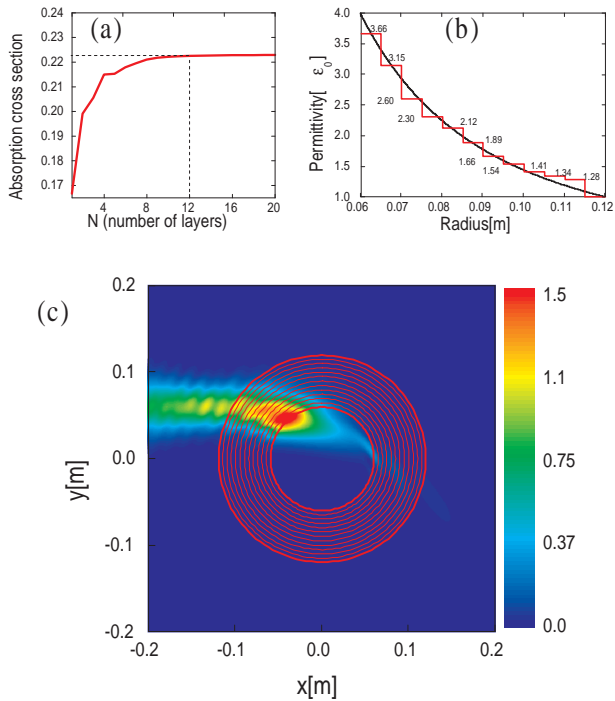}
  \caption{(a). The absorption cross sections varies with different layer numbers in a layered EM black hole. (b). The relative permittivity for each layer (red lines) in a twelve-layer EM black hole. The black curve is
  $\epsilon=(R_2/r)^2$ for comparison.
  (c). The magnetic field intensity pattern near the layered EM black hole, calculated with GMT.}\label{02}
\end{figure}

The above artificial black hole can be designed by using
multi-layered cylindrical structure\cite{JMO}. We break up the
inhomogeneous region $R_1<r<R_2$ into $N$ concentric shells of
isotropic dielectrics with equal thickness. In Fig.\ref{02}(a), we
plot the relationship between the layer number $N$ in $R_1<r<R_2$
and the absorption cross section, which is calculated by
generalized Mie theory (GMT) for multi-layered structure and Eq.
(\ref{acs}). The results show that a twelve-layer structure is
good enough to implement such an EM black hole. As a concrete
example, we will use 12 layers of isotropic materials whose
relative permittivities are shown in Fig.\ref{02}(b). The outmost
layer is set to be air while the permittivity of the inner
absorbing core is $\epsilon_{core}=4+0.33i$ \cite{inabs}.
Fig.\ref{02}(c) shows the magnetic field intensity pattern near
the present layered structure black hole based on GMT. The related
absorption cross section is about 0.2222, which means that the
layered structure works as well as the original one.

\begin{figure}[htbp]
  \centering
  \includegraphics[width=0.45\textwidth]{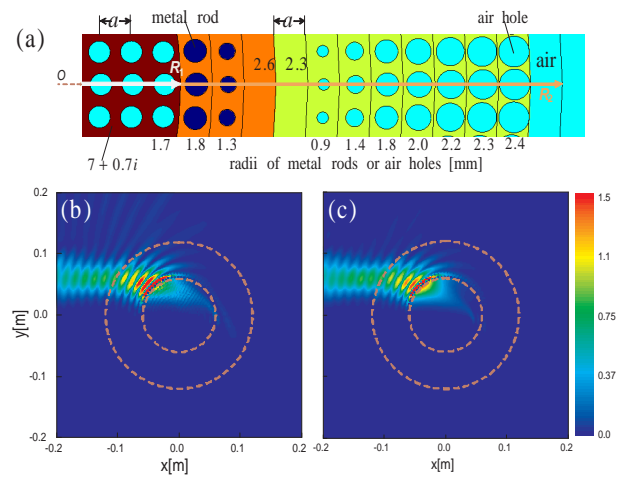}
  \caption{(a). An actual design of the EM black hole with composite materials. The air, Al, PE, PMMA, and PVDF are denoted by blue, dark blue, green, orange, and brown colors respectively. The radii of cylinders in each layer ($r_c$) are also given correspondingly. (b). The magnetic field intensity pattern near the black hole with composite materials described in (a). (c). The same to (b) but replacing the inner absorbing core with PVDF directly. The simulation results are from COMSOL Multiphysics finite element-based electromagnetics solver and the incident beam is a fundamental-mode Gaussian beam for approximations to Eqs. (\ref{inc}) and (\ref{An}).}\label{03}
\end{figure}

However, it would be difficult to find the above twelve layers of
isotropic materials one by one. Based on a two-step approach
\cite{PRL2}, we will implement this layered EM black hole with
only five kinds of real isotropic materials. We will use the
following real materials, air, aluminum (Al) metal rods,
polyethylene (PE), polymethyl methacrylate (PMMA) plexiglass, and
Polyvinylidene fluoride (PVDF). Their related permittivities in
microwave frequencies (near 10 GHz) are about $\epsilon_{PE}=2.3$,
$\epsilon_{PMMA}=2.6$, and $\epsilon_{PVDF}=7+0.7i$ \cite{book}
(Note that $\epsilon_{air}=1$ and $\epsilon_{Al}=-\infty$
\cite{chen}). As the equal thickness of each layer is $5 mm$, we
divide each layer into cells each measuring $5 mm$ in the middle
so that each cell has ``fanlike" shape \cite{chen}. A cylinder
with one kind of materials is placed in the center of each cell
and embedded in a background material. Because the thickness of
each cell/layer is very small when compared with the wavelength,
we can approximate the above twelve layers of isotropic materials
with composite materials of the above five kinds of materials
based on the effective medium theory (EMT)
\cite{hu,jensen,junfeng}. Each layer of isotropic materials can be
approximated by a circular array of cylinders embedded in a
background material, see for details in Fig.\ref{03}(a). From the
EMT, the effective permittivity $\epsilon_{eff}$ of composite
materials with cylinders embedded in a background material with
square lattice satisfies,
\begin{equation}\label{media}
    \frac{\epsilon_{eff}-\epsilon}{\epsilon_{eff}+\epsilon}=\frac{\epsilon_1-\epsilon}{\epsilon_1+\epsilon}f_s,
\end{equation}
where $\epsilon$ is the permittivity of the background material
while $\epsilon_1$ is the permittivity of cylinders with square
lattice, and $f_s$ is the filling ratio of the cylinders. Suppose
the lattice constant is $a=5mm$, $f_s=\pi(\frac{r_c}{a})^2$ with
the radii of cylinders $r_c$.

As the permittivities of the outer nine layers are in the range of
1 and 2.3, we can use air-PE composite materials to approximate
each layer. We shall embed air hole cylinders in PE to approximate
the permittivities from 1.28 to 2.12 (seven layers). The radii of
cylinders in each layer can be obtained from Eq. (\ref{media})
(from $2.4 mm$ to $0.9 mm$, see also in Fig.\ref{03}(a), the PE is
denoted with green color while the air hole cylinders are denoted
with blue color). As permittivities of the inner three layers in
$R_1<r<R_2$ is 3.66, 3.15 and 2.6, we can use Al-PMMA composite
materials to approximate two of them. As the effective
permeability of such composite materials is not unity \cite{hu},
we shall use the square of the effective refractive index (but not
the effective permittivity) to approximate the above two
permittivities (3.66 and 3.15), \emph{i.e.,}
\begin{equation}\label{metal}
    n_{eff}^2=\mu\epsilon_{eff}=\epsilon(1+f_s).
\end{equation}
Likewise, we can obtain the radii of Al cylinders ($1.8 mm$ and
$1.3 mm$ from Eq. (\ref{metal}), see also in Fig.\ref{03}(a), the
PMMA is denoted with orange color while the Al cylinders are
denoted with dark blue color). The inner absorbing core can be
approximated by using air hole cylinders embedded in PVDF. The
inner core is also divided into twelve layers with the same
procedure as that applied to the region $R_1<r<R_2$. The radii of
the air hole cylinders are $1.7 mm$ from Eq. (\ref{media}) (see
also in Fig.\ref{03}(a), the PVDF is denoted with brown color
while the air hole cylinders are denoted with blue color).

Figure \ref{03}(b) shows the magnetic field intensity pattern near
the black hole designed above in Fig.\ref{03}(a). We use COMSOL
Multiphysics finite element-based electromagnetics solver to
perform the simulation, which shows that the black hole with the
above composite materials can approximate both the layered
structure black hole and the original EM black hole very well. In
addition, we found in Fig.\ref{03}(c) that if we replace the inner
core by using PVDF solely (without embedding the air holes), the
device still functions. Hence these two designs both serve for
easy experimental fabrication of EM black holes.

\section{conclusion}
To summarize, we have rigorously studied the properties of an EM
black hole for TM modes. With a two-step approach, we proposed
actual designs of such a black hole by using composite materials
with only five kinds of real isotropic materials, enabling easy
fabrications with nowadays technologies. Our design circumvents
retrieving specific constitute parameters from resonant
structures, the device is expected to function in a broad
bandwidth of frequencies. In particular, due to the simplicities
of such designs, it would be feasible to adapt the same method for
higher frequencies, such as THz, infrared, or even optical
frequencies \cite{val,gab}.

\begin{acknowledgments}
This work was supported by the Soochow University Start-up grant
No. Q4108909, the China 973 program, NNSFC, PCSIRT, MOE of China
(B06011), and the Shanghai Science and Technology Commission. We
thank Dr. Junjie Du, Dr. Zhihong Hang, and Prof. Qiang Cheng for
helpful discussions.
\end{acknowledgments}


\begin{thebibliography}{99}
\bibitem{inabs}
We used $\epsilon_{core}=4+0.33i$ as an example because we found
that such an absorbing material can be implemented with composite
materials of air hole cylinders embedded in Polyvinylidene
fluoride (PVDF) as background material. One can use other values
of $\epsilon^{\prime\prime}$ if some other kinds of absorbing
materials are reqiured.
\end{thebibliography}
\end{document}